\documentclass[sigconf]{acmart}

\copyrightyear{2024}
\acmYear{2024}
\setcopyright{rightsretained}
\acmConference[KDD '24]{Proceedings of the 30th ACM SIGKDD Conference on Knowledge Discovery and Data Mining}{August 25--29, 2024}{Barcelona, Spain}
\acmBooktitle{Proceedings of the 30th ACM SIGKDD Conference on Knowledge Discovery and Data Mining (KDD '24), August 25--29, 2024, Barcelona, Spain}\acmDOI{10.1145/3637528.3671458}
\acmISBN{979-8-4007-0490-1/24/08}

\makeatletter
\gdef\@copyrightpermission{
  \begin{minipage}{0.3\columnwidth}
   \href{https://creativecommons.org/licenses/by/4.0/}{\includegraphics[width=0.90\textwidth]{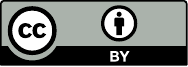}}
  \end{minipage}\hfill
  \begin{minipage}{0.7\columnwidth}
   \href{https://creativecommons.org/licenses/by/4.0/}{This work is licensed under a Creative Commons Attribution International 4.0 License.}
  \end{minipage}
  \vspace{5pt}
}
\makeatother

\usepackage[utf8]{inputenc} 
\usepackage[T1]{fontenc}    
\usepackage{hyperref}       
\usepackage{amsfonts}       
\usepackage{graphicx}
\usepackage{amsthm}
\usepackage{amsmath}
\usepackage{subfigure}
\usepackage{multirow}
\usepackage{enumitem}
\usepackage{bbding}

\newcommand{\eg}{\emph{e.g., }}
\newcommand{\etal}{\emph{et al. }}

\begin{document}


\title{Bias and Unfairness in Information Retrieval Systems: \\ New Challenges in the LLM Era}

\author{Sunhao Dai}
\affiliation{
  \institution{Gaoling School of Artificial Intelligence\\Renmin University of China}
    \city{Beijing}
  \country{China}
  }
\email{sunhaodai@ruc.edu.cn}

\author{Chen Xu}
\affiliation{
  \institution{Gaoling School of Artificial Intelligence\\Renmin University of China}
    \city{Beijing}
  \country{China}
  }
\email{xc_chen@ruc.edu.cn}

\author{Shicheng Xu}
\affiliation{%
  \institution{CAS Key Laboratory of AI Safety\\Institute of Computing Technology Chinese Academy of Sciences}
  \city{Beijing}
  \country{China}
}
\email{xushicheng21s@ict.ac.cn}

\author{Liang Pang}
\authornote{Corresponding author.}
\affiliation{%
  \institution{CAS Key Laboratory of AI Safety\\Institute of Computing Technology Chinese Academy of Sciences}
  \city{Beijing}
  \country{China}
}
\email{pangliang@ict.ac.cn}

\author{Zhenhua Dong}
\affiliation{%
 \institution{Huawei Noah's Ark Lab}
 \city{Shenzhen}
 \country{China}
}
\email{dongzhenhua@huawei.com}

\author{Jun Xu}
\affiliation{
  \institution{Gaoling School of Artificial Intelligence\\Renmin University of China}
    \city{Beijing}
  \country{China}
  }
\email{junxu@ruc.edu.cn}

\renewcommand{\authors}{Sunhao Dai, Chen Xu, Shicheng Xu, Liang Pang, Zhenhua Dong and Jun Xu}
\renewcommand{\shorttitle}{Bias and Unfairness in Information Retrieval Systems: New Challenges in the LLM Era}

\begin{abstract}

With the rapid advancements of large language models (LLMs), information retrieval (IR) systems, such as search engines and recommender systems, have undergone a significant paradigm shift. This evolution, while heralding new opportunities, introduces emerging challenges, particularly in terms of biases and unfairness, which may threaten the information ecosystem. In this paper, we present a comprehensive survey of existing works on emerging and pressing bias and unfairness issues in IR systems when the integration of LLMs. We first unify bias and unfairness issues as distribution mismatch problems, providing a groundwork for categorizing various mitigation strategies through distribution alignment. Subsequently, we systematically delve into the specific bias and unfairness issues arising from three critical stages of LLMs integration into IR systems: data collection, model development, and result evaluation. In doing so, we meticulously review and analyze recent literature, focusing on the definitions, characteristics, and corresponding mitigation strategies associated with these issues. Finally, we identify and highlight some open problems and challenges for future work, aiming to inspire researchers and stakeholders in the IR field and beyond to better understand and mitigate bias and unfairness issues of IR in this LLM era. We also consistently maintain a GitHub repository for the relevant papers and resources in this rising direction at \color{blue}{\url{https://github.com/KID-22/LLM-IR-Bias-Fairness-Survey}}.

\end{abstract}




\maketitle

\section{Introduction}

\begin{figure*}[t]  
    \centering    
    \vspace{-0.2cm}
    \includegraphics[width=0.95\linewidth]{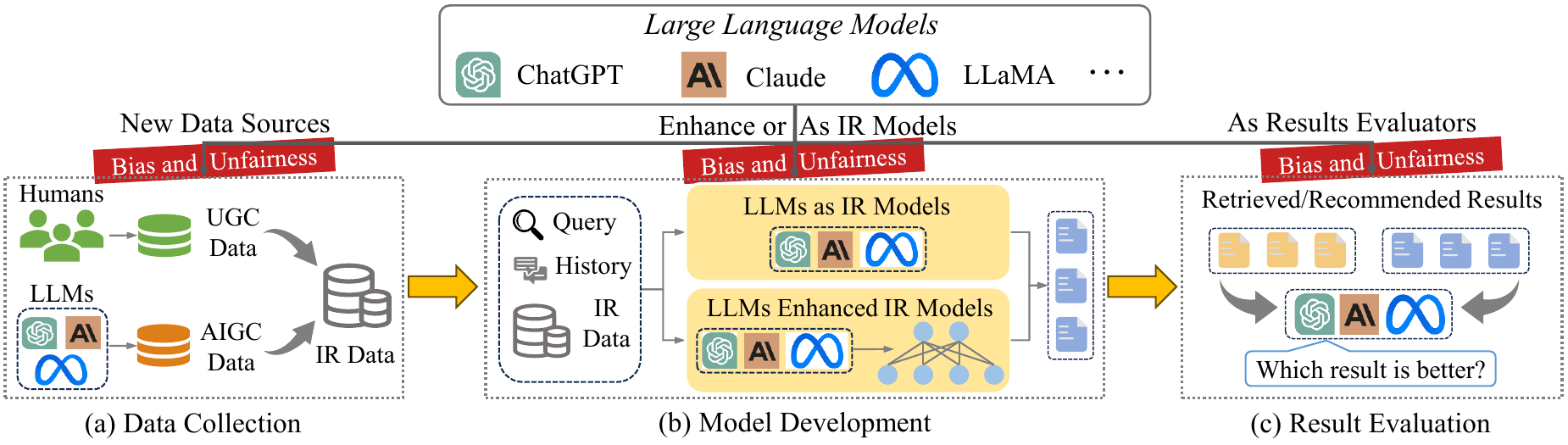}
    \caption{Overview of three stages of the intersection between LLMs and IR systems. (a) LLMs-generated content as new data sources for IR. (b) Incorporating LLMs to enhance or as IR models. (c) Adopting LLMs as results evaluators in IR systems. }
    \label{fig:LLM4IR_overview}  
\end{figure*}

Information Retrieval (IR) systems strive to navigate the era of information overload, facilitating users in acquiring information more efficiently and effectively~\cite{Xu2018IRBook, singhal2001modern, manning2009introduction}. 
The integration of Large Language Models (LLMs) has fundamentally redefined IR systems, including the introduction of LLM-generated data as new IR data sources, the shift from passive retrieval to proactive generation as core paradigms, and adopting LLMs as results evaluators for IR systems~\cite{zhu2023large, ai2023information}. 
These advancements, however, bring forth new challenges in bias and unfairness, affecting the reliability of IR systems and potentially contributing to societal issues like echo chambers~\cite{del2016echo, sharma2024generative} and cognitive interference~\cite{sarason2014cognitive, mihaylov2018dark}. For instance, researchers found that LLMs often retrieve information that deviates from facts and is biased towards LLM-generated content~\cite{dai2023llms, xu2023ai, li2023halueval,chern2023factool}. Moreover, LLMs frequently manifest stereotypes and discriminatory content to users and amplify disparities between items of different socio-economic statuses~\cite{zhang2023chatgpt, xu2023llms, jiang2024item}. 

Recently, much effort has been made around bias and unfairness in the context of LLMs and IR systems. However, the literature is currently fragmented and often lacks a unified definition of these concepts. 
This ambiguity hampers the development of systematic strategies to address these issues effectively. 
To this end, our survey aims to provide a comprehensive and unified perspective that effectively summarizes the emerging challenges and opportunities related to bias and unfairness in the intersection between LLMs and IR systems. Generally, both bias and unfairness issues can be regarded as a \textit{distribution mismatch problem}. Specifically, bias underscores the fact that the predicted information lacks objectivity and truthfulness, highlighting the mismatch with the objective target distribution. Unfairness reveals that the predicted information fails to align with the social values between humans and machines, leading to a mismatch with the subjective target distribution of human values. This perspective not only unifies the nature of these issues but also streamlines the exploration of mitigation strategies.

Our survey begins with a brief overview of how LLMs are integrated into IR systems, setting the stage for understanding the emergence of new bias and unfairness challenges across three pivotal stages of the IR lifecycle: data collection, model development, and result evaluation. Then we propose a unified perspective on bias and unfairness, categorizing them as distribution mismatch problems. Based on this unified view, we categorize mitigation strategies into two principal groups: data sampling, including data augmentation and data filtering, and distribution reconstruction, encompassing rebalancing, regularization, and prompting. Following this taxonomy, we delve into a detailed analysis of several types of bias and unfairness phenomena that arise with the integration of LLMs into IR systems, spanning the aforementioned stages. Our systematic review encompasses a comprehensive examination of these issues and their respective mitigation strategies in recent studies, providing a holistic view of the current landscape and guiding future efforts in eliminating bias and ensuring fairness for more trustworthy IR systems.

\textbf{Difference with Existing Surveys.} 
Several recent surveys have reviewed and discussed the issues of bias and fairness within IR~\cite{chen2023bias, wang2023survey, li2023fairness, zehlike2021fairness, pitoura2022fairness}, primarily focusing on works published before the advent of LLMs. 
With LLMs becoming increasingly prevalent, a new subset of surveys~\cite{gallegos2023bias, xue2023bias, liu2023trustworthy, li2023survey} has paid attention to the bias and fairness challenges presented by LLMs themselves. Additionally, some other recent surveys~\cite{ai2023information, zhu2023large, wu2023survey, lin2023can} have examined how integrating LLMs can enhance and transform traditional IR systems, highlighting some opportunities arising from this integration. 
Compared to these surveys, our work stands apart by offering a comprehensive survey of the emerging and pressing issues of bias and fairness at the intersection of IR and LLMs, employing a novel unified perspective to review the cause and mitigation strategies. 

\textbf{Summary of Contributions.} (1) We provide a novel unified perspective for understanding bias and unfairness as distribution mismatch problems, alongside a detailed review of several types of bias and unfairness arising from integrating LLMs into IR systems. (2) We systematically organize mitigation strategies into two key categories: data sampling and distribution reconstruction, offering a comprehensive roadmap for effectively combating bias and unfairness with state-of-the-art approaches. (3) We identify the current challenges and future directions,  providing insights to facilitate the development of this potential and demanding research area.


\section{A Unified View of Bias and Unfairness}\label{sec:unify}
In this section, we provide a unified view of bias and unfairness from the perspective of distribution alignment and outline the mitigation strategies based on this view.

\subsection{Background}
As shown in \autoref{fig:LLM4IR_overview},  the advent of LLMs has reshaped the whole pipeline of IR systems, typically in the following three stages: data collection, model development, and result evaluation.  

\textbf{LLMs-generated content as new data sources for IR.} The emergence of LLMs has significantly accelerated the growth of Artificial Intelligence Generated Content (AIGC), marking a new era in content creation.
Unlike traditional Professional and User Generated Content (PGC and UGC) sources, AIGC can be produced automatically at scale, potentially dominating the content landscape.~\cite{cao2023comprehensive, wu2023ai, hanley2023machine}.
However, AIGC also reshapes the distribution of the IR data, resulting in new concerns about bias and fairness.

\textbf{Incorporating LLMs to enhance or as IR models.} 
The impressive emergent capabilities of LLMs in understanding, reasoning, and generalization have motivated significant efforts to integrate them into the development of next-generation IR systems~\cite{zhu2023large}.
On one hand, LLMs have been deployed to refine key components of traditional IR systems~\cite{sun2023chatgpt, shi2023replug, wang2023query2doc}, enhancing their effectiveness and efficiency. On the one hand, beyond enhancing existing frameworks, LLMs also introduce a novel paradigm by acting as generative search and recommendation agents~\cite{nakano2021webgpt, gur2023real, dai2023uncovering, zhang2023generative}, directly generating responses to fulfill user queries.

\textbf{Adopting LLMs as results evaluators in IR systems.} 
Human evaluation plays a pivotal role in IR systems, particularly in conversational search and recommendation, such as directly assessing the quality of responses from generative LLM-based IR models~\cite{dai2023uncovering, li2023large, zheng2023generative}.
However, human evaluation comes with significant challenges, including high costs and a lack of reproducibility.
LLMs, with their advanced language modeling and understanding capabilities, offer new possibilities for conducting evaluations of these complex tasks without human evaluation~\cite{chiang2023can, svikhnushina2023approximating, li2024leveraging, gao2024llm}. This shift not only streamlines the evaluation process but also mitigates the substantial costs associated with human evaluations, further facilitating the development of IR systems.

\subsection{Distribution Alignment Perspective}
While the reshaping of the above IR stages by LLMs has introduced numerous new opportunities, it has also given rise to many new and pressing issues related to bias and unfairness. In this section, we formulate the problems of bias and fairness from a distribution alignment perspective, offering a unified framework for understanding these challenges and inspiring mitigation strategies.

Formally, when a user interacts with a typical IR system, he/she may optionally provide his/her personalized information requirement $T$ along with their personalized attributes (typically indicated as user profile $U$) to the system. Subsequently, the goal of the IR system is to retrieve the target information $R$ (\eg documents, items, or advertisements, \etal) for this user, with the user's information requirements and optional interaction history $H$ as the input $Q = \{T, U, H\}$. Let $ \widehat{R} = f(Q)$ be the predicted result either from IR models or directly generated by LLMs, where $f(\cdot)$ is the model. 


Let $P(\widehat{R})$ be the distribution of predicted results for all users, we can unify the bias and unfairness problem as a \textit{distribution mismatch problem} with the ground truth distribution $P(R)$, where $R$ is the target result, which can be defined as follows:
\begin{equation} \label{eq: unified_defintion}
    P(\widehat{R}) \neq P(R).
\end{equation}

Based on the Equation ~\ref{eq: unified_defintion}, the bias and unfairness problem can be explained as follows:

$\bullet$ \textbf{Bias} stems from systemic deviations occurring at various stages of the IR process, from data collection to model design and evaluation. These systemic issues results in the predicted distribution, $P(\widehat{R})$, diverging from the target distribution, $P(R)$, which ideally represents objective and factual realities.

$\bullet$ \textbf{Unfairness} deeply rooted in cultural and societal notions of fairness, aims to align the predicted distribution $P(\widehat{R})$ with a subjective target distribution $P(R)$. It reflects human values and social contracts and evolves with the progress of time.

\begin{figure}[t]  
    \centering    
    \includegraphics[width=0.95\linewidth]{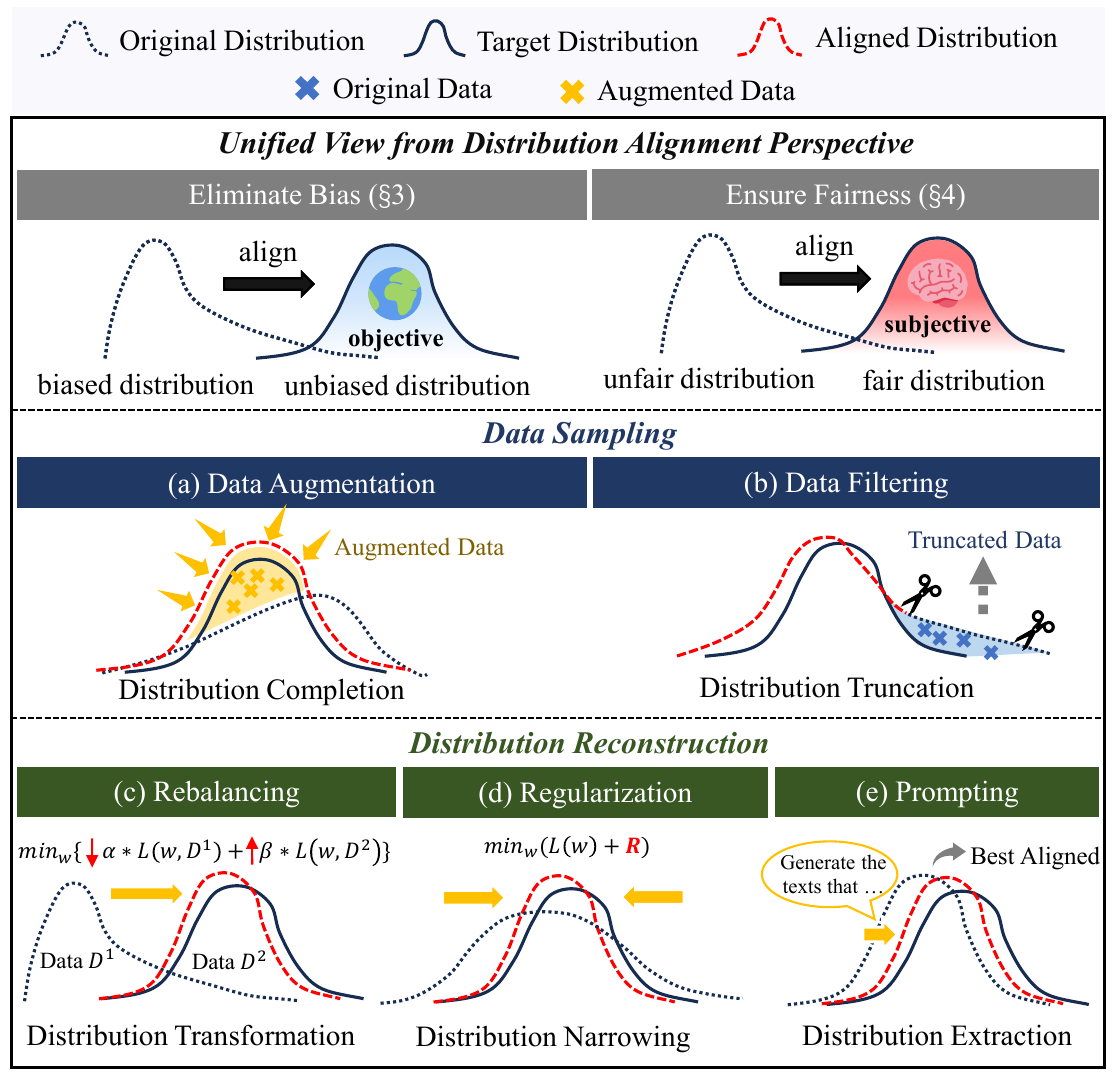}
    \caption{Illustration of different types of mitigation strategies from a unified view of distribution alignment.}
    \label{fig:mitigation_strategies}  
\end{figure}

\subsection{Taxonomies of Mitigation Strategies}
Based on our unified view, we can further systematically categorize mitigation strategies from the view of distribution alignment. Specifically, the goal of mitigation strategies is to align the distribution of retrieved information with a target distribution defined by either objective criteria or subjective social values. As shown in \autoref{fig:mitigation_strategies}, we outline two primary categories and their sub-strategies:

(1) \textbf{Data Sampling} focuses on directly modifying the data:

$\bullet$ \textbf{Data Augmentation} serves as distribution completion, enriching the dataset with additional, often synthetic, data to approximate the target distribution more closely. Techniques such as counterfactual imputation and the incorporation of external knowledge are employed to fill in the gaps in the existing dataset, thus aiming to recover the real distribution more accurately.

$\bullet$ \textbf{Data Filtering} acts as distribution truncation, selecting data subsets that align with the target distribution, ensuring the model's output is representative of desired target outcomes. Techniques like re-ranking and constrained beam search fall under this category, serving as post-processing methods to ensure the retained distribution segment matches the target distribution.

(2) \textbf{Distribution Reconstruction} aims at adjusting the predicted distribution:

$\bullet$ \textbf{Rebalancing} transforms the predicted distribution through techniques like reweighting or resampling, to reflect the target distribution more accurately. Common strategies include adjusting the loss weights for various groups to achieve equilibrium, thereby realigning the predicted distribution with the target distribution.

$\bullet$ \textbf{Regularization} narrows the predicted distribution by introducing constraints that encourage the model to learn the target distribution more faithfully. It encompasses both implicit approaches, such as adversarial learning, and explicit ones, like regularization techniques, to modify the distribution directly.

$\bullet$ \textbf{Prompting} extracts the best aligned distribution by directly employing specific prompts. This approach guides LLMs to generate outputs more likely from the target distribution, facilitating an alignment with the desired target distribution.

Through the lens of distribution alignment, these strategies offer a structured and coherent approach to eliminate bias and ensure fairness of IR systems in IR systems. 
In the following sections, we will conduct a detailed review of various emerging issues related to bias and unfairness, their mitigation solutions, and their integration within the distribution alignment framework.

\section{Cause and Mitigation of Bias}\label{sec:bias}
As summarized in \autoref{tab:bias_type}, we present an in-depth review of different types of bias at the different stages of the intersection between LLMs and IR systems and discuss the corresponding mitigation strategies.

\begin{table*}[t]
\small
\caption{The taxonomy of different types of bias in the intersection between LLMs and IR systems.}
\vspace{-0.2cm}
\label{tab:bias_type}
\centering
\resizebox{1\linewidth}{!}
{
\begin{tabular}{c|c|c|c|c|c|c}
\bottomrule
\multirow{3}{*}{Sourced Stage}            & \multirow{3}{*}{Type}                  & \multicolumn{5}{c}{Mitigation Strategies}           \\
\cline{3-7}
& & \multicolumn{2}{c|}{Data Sampling} & \multicolumn{3}{c}{Distribution Reconstruction} \\
\cline{3-7}
                  &                       & Data Augmentation    & Data Filtering & Rebalancing & Regularization & Prompting \\
\hline
\multirow{2}{*}{Data Collection}   & Source Bias           &               &   \cite{chen2024spiral}        &             &   \cite{dai2023llms, xu2023ai, zhou2024source}             &           \\
\cline{2-7}
                  & Factuality Bias         &  \cite{ram2023context, yu2023improving, xu2023search,press2022measuring,xu2024unsupervised,xu2024list,gunasekar2023textbooks}                     &    \cite{gunasekar2023textbooks, touvron2023llama,yang2022unified}       &     &     &   \cite{xu2023search,press2022measuring,wang2023selfconsistency,sun2022recitation}     \\
\hline
\multirow{4}{*}{Model Development}  & Position Bias         &  \cite{hou2024llmrank, qin2023large, tang2023found, wu2023exploring, luo2023recranker, zhang2023agentcf}                     &           &   \cite{ma2023large, wu2023exploring}          &                &  \cite{hou2024llmrank}         \\
\cline{2-7}
                  & Popularity Bias       &  \cite{wang2023improving, zhang2023agentcf}                     &           &             &                &  \cite{hou2024llmrank, deldjoo2024understanding, spurlock2024chatgpt}         \\
\cline{2-7}
                  & Instruction-Hallucination Bias         & \cite{mishra2021cross,sanh2021multitask,wang2022self}                     &           &             &    \cite{fernandes2023bridging}           &   \cite{phillips2012generalized,ye2023investigating}    \\ \cline{2-7}
                            & Context-Hallucination Bias         &   \cite{bai2024longalign,fu2024data}                   &           &             &               &      \\
\hline
\multirow{3}{*}{Result Evaluation} & Selection Bias        &   \cite{li2023split, zheng2023generative, pezeshkpour2023large, chern2024can, zheng2024judging, li2023prd, wang2023large, chu2024pre}                    &           &  \cite{zheng2023large, wang2023large, liusie2024teacher}           &                &   \cite{ zheng2024judging, pezeshkpour2023large, wang2023large, kim2023evallm}        \\
\cline{2-7}
                  & Style Bias            &                       &           &            &                &  \cite{wu2023style, zheng2024judging}         \\
\cline{2-7}
                  & Egocentric Bias &  \cite{li2023prd}                     &           & \cite{liu2023gpteval}            &                & \cite{liu2023gpteval, hasanbeig2023allure}          \\
\toprule
\end{tabular}
}
\end{table*}


\subsection{Bias in Data Collection}

In this subsection, we categorize the bias caused by data collection into two groups: source bias and factuality bias.

\subsubsection{Source Bias} Source bias emerges when the incorporation of LLM-generated content into the corpus of the IR systems:


$\bullet$ \textbf{Definition.} Information retrieval models tend to rank content generated by LLMs higher than content authored by humans.

Specifically, as LLMs fuel the rapid expansion of LLM-generated content, the corpus for IR systems now increasingly encompasses a mix of human-written and LLM-generated texts. 
Recent studies~\cite{dai2023llms, xu2023ai, dai2024cocktail} highlight that modern retrieval models, especially those leveraging neural matching techniques, tend to favor LLM-generated content over human-authored content with similar semantics. This preference stems from the unique representations embedded in LLM-generated content, which neural retrieval models can capture and thus assign a higher estimated relevancy score~\cite{dai2023llms, xu2023ai}. More severely, this source bias is further amplified when LLM-generated content is included in model training, presenting a challenge to the information ecosystem~\cite {xu2023ai}. \citet{zhou2024source} further explores this escalation of source bias in user, data, and recommender systems feedback loop. \citet{tan2024blinded} further that this bias will extend from retrievers to the readers and \citet{chen2024spiral} uncover this bias in the RAG systems.

To counteract source bias, recent studies have introduced debiased constraints into the training objectives of IR retrieval models~\cite{dai2023llms, xu2023ai, zhou2024source}. This strategy aims to correct the skewed relevancy predictions favoring LLM-generated content, ensuring fair treatment between human-written and LLM-generated content. 
By adopting a distribution alignment perspective, such mitigation efforts strive to recalibrate the IR models' relevancy distribution towards an ideal state, where documents are judged equally based on their semantic content rather than their source. 

\subsubsection{Factuality Bias}  
As AIGC increasingly becomes a part of the data sources for IR systems, it inevitably introduces a significant amount of non-factual or ``hallucinated'' content. This introduction alters the distribution of IR system data, thereby leading to biases in the retrieval process.


$\bullet$ \textbf{Definition.} LLMs may produce content that does not align with recognized factual information of the real world.

Many studies have shown that LLMs are at risk of generating factual errors. For instance, TruthfulQA~\cite{lin2021truthfulqa} has highlighted that language models generate many false answers that mimic popular misconceptions in question-answering tasks and have the potential to deceive humans. Besides, merely scaling up models is not promising for improving truthfulness, which means that LLMs still face challenges in generating factually correct content. \citet{lee2022factuality} show LLMs are susceptible to generating text with nonfactual in an open-ended generation because of the ``uniform randomness'' at every sampling step. FActScore~\cite{min2023factscore} finds that LLMs lag significantly behind humans in ensuring the factual consistency of long-form text generation. 
Other studies~\cite {mckenna2023sources} reveal that LLMs also exhibit factual hallucination in natural language inference tasks. In addition to the above studies, many large-scale benchmarks~\cite{li2023halueval,chern2023factool,chen2023complex} also indicate that LLMs exhibit factuality bias in multi-task and multi-domain scenarios.

Previous studies find that the flawed data source and inferior data utilization are two important causes of factuality bias~\cite{huang2023survey}. Specifically, some low-quality, factual errors, and long-distance repetition in the training texts harm the factual correctness of the text generated by LLMs~\cite{lin2021truthfulqa,lee2021deduplicating,bender2021dangers}. The coverage of knowledge by training data also limits the correctness of LLMs in generating knowledge in some rare or specialized fields~\cite{singhal2023towards, onoe2022entity, katz2024gpt}. In addition to the training data, LLMs usually resort to shortcuts to generate the texts depending on position close and co-occurred words rather than understand the knowledge itself~\cite{li2022pre,kandpal2023large,kang2023impact} and always fail to recall the knowledge that has been memorized~\cite{mallen2022not,zheng2023does}.

To mitigate factuality bias, in the training of LLMs, some methods focus on providing high-quality and factually correct training data for LLMs~\cite{gunasekar2023textbooks, touvron2023llama}. In the inference, previous studies can be divided into two categories. One is with the help of an external factual knowledge base such as retrieval-augmented generation~\cite{ram2023context, yu2023improving, xu2023search,press2022measuring,xu2024unsupervised,xu2024list,gao2022rarr}. The other is to improve the ability of LLMs themselves such as Self-Consistency~\cite{wang2023selfconsistency} and Dola~\cite{chuang2023dola}.

\subsection{Bias in Model Development}
Incorporating LLMs into IR models introduces inherent randomness in the generation results, potentially leading to inconsistent outcomes. In this subsection, we categorize the bias in model development into four groups: position bias, popularity bias, input-hallucination bias, and context-hallucination bias.

\subsubsection{Position Bias} Position bias emerges notably in scenarios where LLMs are utilized directly as retrieval or recommender systems~\cite{tang2023found, hou2024llmrank}, characterized by the preference of documents or items based on their input positions:


$\bullet$ \textbf{Definition.} 
LLM-based IR models tend to give preference to documents or items from specific input positions.

Recent works have highlighted that the order of candidate documents or items can significantly impact the performance of LLM-based IR models, while conventional IR models are often not affected by the changing of input orders~\cite{xu2024prompting, hou2024llmrank, tang2023found}. For instance, LLM-based models often show a preference for content positioned at the beginning or end of a list, neglecting the contributions of items in the middle. This ``lost in the middle'' suggests that these LLM-based models may not fully utilize the context provided by items that don't occupy prominent positions in the input sequence~\cite{liu2024lost}.

There are a number of works on mitigating position bias, and we can categorize them into three lines based on our distribution alignment framework. (1) Prompting: This approach involves carefully designed prompts to encourage the models to disregard the input's order~\cite{hou2024llmrank, dai2023uncovering}. Nonetheless, due to LLMs' prompt sensitivity, this requires precise and task-specific prompt engineering across various tasks and domains.
(2) Data Augmentation: This approach has been explored in numerous studies, which is a form of data augmentation that involves random shuffling of candidates followed by aggregation to determine the final ranking. \cite{hou2024llmrank, qin2023large, tang2023found, wu2023exploring, luo2023recranker, zhang2023agentcf}. For instance, 
\citet{tang2023found} introduced a permutation self-consistency method, offering theoretical guarantees under certain conditions, enabling models to produce and aggregate potential outcomes from various candidate permutations, enhancing result stability. (3) Rebalancing: This method counteracts position bias by adjusting the prior distribution sensitive to positions. It recalibrates the model's output, addressing the inherent bias towards item positions~\cite{ma2023large, wu2023exploring}.

While these strategies offer pathways to counteract position bias, they present challenges, notably the increased computational demand associated with processing multiple permutations~\cite{qin2023large, hou2024llmrank, tang2023found}. Future work should aim to balance effectiveness with efficiency to develop more stable and unbiased LLM-based IR systems.

\subsubsection{Popularity Bias} Popularity bias has been a widely studies issue in traditional IR models, with extensive research highlighting its effects on the so-called ``Matthew effect'' issue~\cite{abdollahpouri2019unfairness, zhang2021causal}. This bias is characterized by the long-tail phenomenon, where a minority of popular items garners a majority of user interactions. Consequently, models trained on such skewed data tend to favor these popular items, often over-representing them in results and further exacerbating their dominance~\cite{chen2023bias, zhu2021popularity}. However, the advent of LLM-based IR models introduces new dimensions to the challenge of popularity bias, which can be defined as follows:


$\bullet$ \textbf{Definition.} 
LLM-based IR models tend to prioritize candidate documents or items with high popularity levels.
 
Unlike conventional models, LLM-based IR models do not merely reflect the popularity distributions of the target finetuning dataset. They are also inclined to retrieve or recommend items popular in the pre-training corpora of the LLMs~\cite{he2023large, bao2023bi}. For instance, the vast training data of LLMs, encompassing a wide array of content, means that certain documents or items may be over-represented, influencing the model to prefer these familiar documents or items in retrieval and recommendation tasks. As a result, LLM-based IR models may not only exhibit the existing popularity bias found within the target finetuning dataset but also introduce a new bias based on the content's prevalence in their pre-training data. This extension of popularity bias in LLM-based IR models presents a more complex problem.

To combat popularity bias in LLM-based IR models, existing methods explore two main solutions: (1) Data Augmentation: \citet{wang2023improving} proposes two data augmentation strategies to diversify the dataset by incorporating more underrepresented content, aiming to balance the final recommendation results.
(2) Prompting: Alternative methods involve crafting specific instructions to directly intervene in the LLM's output, such as encouraging an equitable mix of popular and long-tailed items in results.
~\cite{hou2024llmrank, deldjoo2024understanding, spurlock2024chatgpt}
However, addressing this expanded notion of popularity bias in LLM-based IR systems requires new strategies in future work that account for both the inherent biases of the training data and the additional biases introduced by the LLMs' pre-training corpora. 

\subsubsection{Instruction-Hallucination Bias} Instruction-hallucination bias emerges when LLMs are used as retrievers, rerankers, or recommenders but do not fully follow the user's instructions:


$\bullet$ \textbf{Definition.} Content generated by LLM-based IR models may deviate from the instructions provided by users.

Recent studies reveal that LLMs often struggle to adhere fully to users' instructions across various natural language processing tasks,  such as dialogue generation~\cite{dziri2021evaluating}, question answering~\cite{durmus2020feqa} and summarization~\cite{maynez2020faithfulness,pu2023summarization}.
These instructions comprise the users' intent or input task (e.g., reranking a document list) and the specific content or object (e.g., the document list) targeted by the task. Deviations from the input task suggest that LLMs may misinterpret the tasks users intend to execute. For instance, when deployed as recommenders, LLMs might not grasp users' requests for items with particular characteristics, leading to recommendations that do not match the request~\cite{fan2023recommender}. Similarly, contradictions with the input content reveal that in tasks like reranking, LLMs may produce results that are inconsistent with, or even absent from, the given instructions, showcasing a gap in understanding and fulfilling the specified requirements~\cite{zhu2023large}.

The key to mitigating the instruction-hallucination bias is to enhance the instruction following the ability of large language models. For example, some works propose high-quality instruction fine-tuning datasets such as Natural Instructions~\cite{mishra2021cross}, Public Pool of Prompts~\cite{sanh2021multitask}, and Self-Instruct~\cite{wang2022self}, etc. Besides, other studies try to further align content generated by LLMs with human preferences via reinforcement learning from human feedback~\cite{fernandes2023bridging}.

\subsubsection{Context-Hallucination Bias} This bias emerges when LLMs are used as recommenders or re-rankers in scenarios with long and rich context, which can be defined as:


$\bullet$ \textbf{Definition.} LLMs-based IR models may generate content that is inconsistent with the context.

There have been many studies showing that LLMs run the risk of generating content that is inconsistent with the context, especially in scenarios where the context is very long and multi-turn responses are needed~\cite{liu2024lost,qiu2024clongeval,bai2024longalign,li2023loogle}. When using LLMs as recommenders in scenarios with complex context such as multiple rounds of conversation history and user portrait information, LLMs may give the items that are contradictory with the conversation history and user preferences. This emerges when LLMs cannot understand the context or fail to maintain consistency throughout the conversation~\cite{zhang2023sirens}, which is mainly because LLMs still have limitations in processing long texts~\cite{liu2024lost}. Therefore, the main research to mitigate the context-hallucination bias focuses on improving the memory and processing capabilities of LLMs for long texts~\cite{xu2023retrieval,chen2023longlora,wang2024augmenting}.

\subsection{Bias in Result Evaluation}
When adopting LLMs as result evaluators in IR systems, the following three types of bias emerge, including selection bias, style bias, and egocentric bias.

\subsubsection{Selection Bias} A primary challenge in utilizing LLMs as evaluators is that they are sensitive to the order/ID tokens of candidate responses, a phenomenon known as selection bias:


$\bullet$ \textbf{Definition.} 
LLM-based evaluators may favor the responses at specific positions or with specific ID tokens.

For example, \citet{zheng2023large} have demonstrated that gpt-3.5-turbo  exhibits a preference for choice ``C'', while llama-30B shows a preference for choice ``A'' across various benchmarks.  Other works have revealed a common tendency among LLMs to favor responses positioned at the first position ~\cite{wang2023large, pezeshkpour2023large, koo2023benchmarking, zheng2024judging, chen2024humans}. Selection bias may stem from an imbalance in how answers of different positions or with distinct ID options are represented in the training data~\cite{zheng2024judging, zheng2023large}. Further investigation reveals that this bias tends to amplify when LLM-based evaluators are uncertain about the prediction between the top-ranked choices\cite{wang2023large, zheng2023large}.

To address selection bias, several approaches have been explored: (1) Prompting: Simple prompt-based methods, such as incorporating few-shot examples or employing chain-of-thought and reference-guided judgment, have been proposed \cite{zheng2024judging, pezeshkpour2023large, kim2023evallm}.  (2) Data Augmentation: Strategies like position or token switching aim to eliminate selection bias by diversifying the evaluation context~\cite{zheng2024judging, chern2024can, zheng2023generative, wang2023large, pezeshkpour2023large}. While these methods can enhance the robustness of evaluations, they are often time-consuming and costly. (3) Rebalancing: \citet{zheng2023large} utilized probability decomposition techniques to estimate the prior distribution of specific positions or tokens associated with responses, which helps in aligning evaluations closer to objective standards.
\citet{wang2023large} have developed a calibration framework that integrates Human-in-the-Loop to calculate balanced position diversity entropy for final selection. Despite the availability of these strategies, effectively mitigating selection bias in LLM-based evaluators remains a challenge, requiring further research for more efficient solutions.

\subsubsection{Style Bias} Style bias can be viewed as a form of aesthetic bias, where the appeal of presentation overshadows the substance, leading to a preference for responses that, while polished in appearance, might harbor factual inaccuracies:



$\bullet$ \textbf{Definition.} 
LLM-based evaluators may favor the responses with specific styles (e.g., longer responses).

For instance, several studies have identified a clear preference in LLMs for longer responses over shorter ones, emphasizing form over the actual quality of the content~\cite{wu2023style, zheng2024judging, liu2023llms, saito2023verbosity}. Moreover,  \citet{chen2024humans} have observed an inclination towards content with visually engaging elements, such as emojis or references, even when such content may include factual errors. \citet{huang2024empirical} suggest that this bias may stem from the training process of LLMs, which emphasizes generating fluent and verbose responses, thereby inadvertently leading them to prefer these characteristics when employed as evaluators.

Addressing style bias remains challenging, with current strategies mainly counteract overemphasis on stylistic features through prompts. However, these measures are often insufficient, highlighting the need for modifications in LLM architecture and training approaches to mitigate this bias effectively in future work. 

\subsubsection{Egocentric Bias} With LLMs being extensively utilized in the development of IR models, egocentric bias has emerged as a new bias during the automated evaluation conducted by LLMs~\cite{liu2023llms, liu2023gpteval, koo2023benchmarking, xu2024perils}, which can be defined as follows:


$\bullet$ \textbf{Definition.} 
LLM-based evaluators prefer the responses generated by themselves or LLMs from the same family.

A recent work~\cite{liu2023llms} has identified that language model-driven evaluation metrics, such as BARTScore~\cite{yuan2021bartscore}, T5Score~\cite{qin2022t5score}, and GPTScore~\cite{fu2023gptscore}, inherently favor texts produced by their underlying LMs, especially in summarization tasks. 
\citet{liu2023gpteval} and \citet{zheng2024judging} further highlighted that when acting as evaluators, LLMs demonstrate a clear bias towards outputs generated by themselves over those from other models or human contributors. This bias could stem from that the LLM may share the same for both the model development phase and result evaluation phase~\cite{liu2023gpteval}.

The emergence of egocentric bias introduces the risk of self-reinforcement for LLMs, particularly when they are further trained using rewards from LLM-based evaluations. 
This scenario can lead to LLMs overfitting to their own evaluation criteria, intensifying self-preference in next-generation model~\cite{liu2023gpteval}.
Current strategies for mitigating egocentric bias primarily involve employing diverse LLMs as evaluators to foster peer discussions~\cite{li2023prd}, thereby reducing the preference for any specific LLM and enhancing the robustness of evaluation outcomes. However, this strategy inevitably increases the evaluation costs. Future research must explore more efficient solutions to ensure fair and unbiased evaluation.



\begin{table*}[t]
\caption{The taxonomy of different types of unfairness in the intersection between LLMs and IR systems.}
\label{tab:fairness_type}
\vspace{-0.5em}
\centering
\resizebox{1\linewidth}{!}
{
\begin{tabular}{c|c|c|c|c|c|c}
\bottomrule
\multirow{3}{*}{Sourced Stage}            & \multirow{3}{*}{Type}                  & \multicolumn{5}{c}{Mitigation Strategies}           \\
\cline{3-7}
& & \multicolumn{2}{c|}{Data Sampling} & \multicolumn{3}{c}{Distribution Reconstruction} \\
\cline{3-7}
                  &                       & Data Augmentation    & Data Filtering & Rebalancing & Regularization & Prompting \\
\hline
\multirow{2}{*}{Data Collection}   & User Unfairness           &   \cite{ghanbarzadeh2023gender, zhang2023chatgpt, xu2023llms, lu2020gender, sun2022moraldial, ung2021saferdialogues}            &    \cite{raffel2020exploring, ngo2021mitigating}     &   \cite{orgad2022blind, deldjoo2024cfairllm} &        \cite{qian2019reducing,bordia2019identifying,huang2019reducing}        &     \cite{fang2024bias}      \\
\cline{2-7}
                  & Item Unfairness         &     \cite{zou2022automatic, rathod-etal-2022-educational}                  &   \cite{guenole2024pseudo}        &   \cite{jiang2024item}   &                 &     \cite{laverghetta2023generating, fang2024bias}      \\
\hline
\multirow{2}{*}{Model Development}  & User Unfairness           &     \cite{wang2021dynamically}          &   \cite{wang2021dynamically, meade2023using, schramowski2022large, shuster2022blenderbot}       &    \cite{han2021balancing, zayed2023should}          &       \cite{liu2019does, woo2023compensatory, park2023never, wang2023toward, garimella2021he, zhou2023causal, ouyang2022training, bai2022constitutional}         &    \cite{deldjoo2024cfairllm, zhang2023chatgpt, yang2023adept, hua2023up5}       \\
\cline{2-7}
                  & Item Unfairness         &      \cite{zu2023automated}                 &   \cite{chung2023increasing, kim2022critic}       &  \cite{jiang2024item}  &   \cite{friedrich2023fair}             &    \cite{zu2023automated, li2023preliminary, deldjoo2024understanding}      \\
\hline
\multirow{2}{*}{Result Evaluation} & User Unfairness           &   \cite{karra2023estimating}            &      \cite{li2023tailoring}     &              &                 &     \cite{safdari2023personality, pan2023llms, yang2023psycot, jiang2024evaluating, bai2023fairmonitor}      \\
\cline{2-7}
                  & Item Unfairness         &      \cite{grosse2023studying}                 &             &   \cite{shi2023detecting, akyurek2022towards}  &                  &    \cite{wang2023recagent, zhang2023agentcf, zhang2023generative, wang2023wasa, sander2024watermarking}       \\
\toprule
\end{tabular}
}
\end{table*}

\section{Cause and Mitigation of Unfairness}\label{sec:fairness}
As summarized in \autoref{tab:fairness_type}, we will review the cause and the mitigation strategies for the unfairness problem of IR in the LLM era.

\subsection{Fairness Concepts}

Sociological researches acknowledge multiple cultural variations in perceptions of fairness~\cite{tyler2002procedural,tyler1995social}. In IR systems, achieving fairness often entails ensuring that the retrieved documents or recommended items align with cultural values~\cite{mehrabi2021survey, li2023fairness}, including principles such as gender equality~\cite{fang2024bias, zhang2023chatgpt, xu2023llms, li2023fairness}, addressing disadvantages~\cite{xu2023p, patro2020fairrec}, and avoiding discriminatory language~\cite{wen2023unveiling, gallegos2023bias}.

Researchers have revealed that various multi-stakeholders involved in IR systems~\cite{abdollahpouri2020multistakeholder, chen2023bias}, such as users and items, often have distinct perspectives on fairness considerations. In IR, user fairness and item fairness are often associated with two sociological concepts: equality and distributive justice~\cite{xu2023p}.

$\bullet$ \textbf{User Fairness.}
Everyone should be treated the same and provided the same resources to succeed. This implies that the IR systems should deliver equitable and non-discriminatory information services to different users.

$\bullet$ \textbf{Item Fairness.}
The resources should be equally distributed based on needs. This implies that the IR systems should afford more opportunities (\eg exposures) to weaker items, striving to equalize the opportunities across diverse items.

\subsection{Unfairness in Data Collection}
In this section, we will elucidate the underlying causes of unfairness in the data collection process and subsequently outline current mitigation strategies to address these issues.

\subsubsection{User Unfairness}
In the context of user unfairness, one primary cause stems from the existing taxonomic, discriminatory, and offensive content in the training data, disproportionately affecting specific groups. The inclusion of these contents within the existing material can be attributed to historical and cultural reasons~\cite{ntoutsi2020bias, zhuo2023exploring, beukeboom2019stereotypes}, or they may be generated by LLMs~\cite{fang2024bias}. 
For example, when training LLMs, it is common to encounter discriminatory content~\cite{abid2021persistent, fang2024bias, das2024under}.
Discrimination against certain groups can also stem from unbalanced data collection, where the insufficient representation of diverse perspectives leads to unfair outcomes~\cite{ghanbarzadeh2023gender, mehrabi2021survey}. 
The presence of unbalanced data can contribute to the perpetuation of historical and cultural stereotypes~\cite{ellemers2018gender, haines2016times, ladhak2023pre} or systematic influences~\cite{chen2023bias}. 


Previous works have employed various methods to mitigate unfairness during data collection by redistributing the existing document corpus. Specifically, some work~\cite{ghanbarzadeh2023gender,zhang2023chatgpt, xu2023llms, lu2020gender} create matched pairs (\eg male or female) to ensure a more equitable dataset and other methods~\cite{dixon2018measuring, sun2022moraldial, ung2021saferdialogues} add non-toxic examples for groups. Other approaches~\cite{orgad2022blind, deldjoo2024cfairllm} suggest using down-weighting samples containing social group or discriminated information. Moreover, other studies~\cite{raffel2020exploring, ngo2021mitigating} propose to filter out discriminated or taxonomic content from web-scale datasets. Finally, instruction fine-tuning or RLHF has also been shown to be effective in promoting fairness~\cite{touvron2023llama,lu2020gender}.

\subsubsection{Item Unfairness}
For item unfairness, one primary cause is likely unbalanced data collection, where the insufficient representation of certain items leads to disparities in the IR process~\cite{jiang2024item}. 
Another reason raised is that LLMs cannot only retrieve existing items but also generate new items, contributing to the potential introduction of novel content and perspectives~\cite{das2024under, laverghetta2023generating, guenole2024pseudo, gotz2023let, zu2023automated}. However,  these newly generated items may still encounter discrimination issues~\cite{laverghetta2023generating, guenole2024pseudo}.

To mitigate item unfairness during data collection, several studies have developed methods to generate non-discriminatory items. For instance, \citet{zou2022automatic} and \citet{rathod-etal-2022-educational} suggest using specific templates to enrich training data with a variety of safe and equitable question-answer pairs, thereby improving LLM-based IR models' training.   \citet{guenole2024pseudo} introduce pseudo-item discrimination techniques for filtering out non-discriminated items. Additionally, other studies~\cite{laverghetta2023generating, fang2024bias} advocate for employing fairness-aware prompts to produce newly non-discriminatory items. Furthermore, ~\citet{jiang2024item} proposes to re-weight different item samples to enhance item fairness.

\subsection{Unfairness in Model Development}
In this section, we will analyze the causes of unfairness in the model development phase and explore mitigating strategies to address and minimize these disparities.

\subsubsection{User Unfairness}

When adapting LLMs as information retrievers, researchers have observed that the extensive knowledge gained during pre-training may introduce risks of user unfairness~\cite{xu2023llms, zhang2023chatgpt, ghanbarzadeh2023gender, deldjoo2024cfairllm}, highlighting the need for careful consideration and mitigation strategies in deploying such models. Studies~\cite{zhang2023chatgpt, deldjoo2024cfairllm, ghanbarzadeh2023gender}  have shown that utilizing explicit user-sensitive attributes like gender or race in LLMs may lead to the generation of discriminated recommendation results or unfair answers to specific questions. Moreover, it has been observed that LLMs can learn implicit attributes, such as user names and email addresses, and utilize them to generate discriminated content~\cite{xu2023llms, wolfe2021low}. 

To address user unfairness in the model development phase, prior research has suggested mitigating unfairness during the fine-tuning process. Some studies~\cite{deldjoo2024cfairllm} investigate how different intersectional prompts affect recommendation fairness and UP5~\cite{hua2023up5, yang2023adept} propose to conduct prompt tuning to get an effective fair-aware prompt. \citet{han2021balancing, zayed2023should} proposes to set different weights of loss for different samples containing discriminated content. Meanwhile, various studies~\cite{liu2019does, woo2023compensatory, park2023never, wang2023toward, garimella2021he, zhou2023causal} explore the incorporation of a fairness-aware regularizer to assist LLMs-based models in generating more equitable content. \citet{wang2021dynamically} proposes to remove unfair information from LLMs-based embeddings by generating adversarial examples. There are also some works~\cite{wang2021dynamically, meade2023using, schramowski2022large, shuster2022blenderbot} that recommend employing a filtering-list approach or comparing model outputs with safe samples to proactively prevent the inclusion of discriminatory words and enhance the fairness of the generated content.


\subsubsection{Item Unfairness}

In the realm of item unfairness, previous studies have revealed that LLMs-based recommendation models are more prone to generating unfair outcomes for items when compared to traditional models~\cite{jiang2024item}. Moreover, certain works~\cite{Zarifhonarvar2023chatgpt4job} have also found that LLMs will also recommend unfair job opportunities to users. The embedding of item unfairness in the model development process can contribute to increased item polarization through reinforcement, potentially creating echo chambers that limit users' exposure to diverse perspectives~\cite{del2016echo}.

Efforts to address item unfairness encompass a range of strategies: \citet{zu2023automated} have employed prompt-based learning to train GPT-2, leveraging this approach to generate distractors for fill-in-the-blank vocabulary items; some studies~\cite{chung2023increasing, kim2022critic} propose to utilize decoding strategies to decrease the probability of existing tokens/items; and \citet{friedrich2023fair} suggests integrating a fairness term into the LLMs-based diffusion process to introduce a shifting fairness consideration, aimed at generating new items that are less likely to contain discriminatory elements. Moreover, ~\citet{jiang2024item} advocates for the re-weighting of different items to effectively mitigate unfairness in recommendation tasks. Other studies~\cite{li2023preliminary, deldjoo2024understanding} propose some prompt-aware methods to mitigate provider fairness.

\subsection{Unfairness in Result Evaluation}
To assess the fairness performance of IR models effectively, it is essential to measure the distributions of social values, which represent the fairness objectives inherent in the evaluation process~\cite{chang2023survey}. However, human evaluation demands significant labor resources. Therefore, recently, LLMs have been employed to simulate human or real systems to facilitate evaluation processes efficiently.

\subsubsection{User Unfairness}
Similarly, user unfairness typically arises when LLMs-based evaluators fail to accurately simulate the behaviors exhibited by real humans. For example, ~\citet{zhang2023safetybench} propose leveraging psychological knowledge to assess the simulated human ability of LLMs, but they discover that LLMs frequently exhibit certain group behavior towards certain human groups. 

To enhance the capability of LLMs as fair evaluators for IR systems, previous studies have devised several methods.
Approaches include designing or learning specific prompts informed by psychological insights to better simulate diverse human groups ~\cite{safdari2023personality, pan2023llms, yang2023psycot, jiang2024evaluating}, augmenting training data with additional human personality information to refine LLMs as evaluators~\cite{karra2023estimating}, and adopting innovative techniques like the unsupervised constructed personalized lexicon (UBPL) to manipulate their individual characteristics~\cite{li2023tailoring}. Furthermore, \citet{bai2023fairmonitor} proposes four stages to identifying discrimination patterns in queries.

\subsubsection{Item Unfairness}
In evaluating item unfairness, when turning from discriminant style to generation style,  a primary concern arises from attributing credit to the generated items~\cite{shi2023detecting, grosse2023studying, akyurek2022towards, wang2023wasa, sander2024watermarking}, as achieving item fairness necessitates tracing this credit back to the item provider for a comprehensive assessment.

To tackle these challenges, several studies \cite{wang2023recagent, zhang2023agentcf, zhang2023generative} have developed IR agents that use world knowledge to fairly distribute item credits, aiming to reduce biases and enhance fairness. \citet{shi2023detecting} employs the MIN-K\% Prob technique to check whether an item exists in the training data. Meanwhile, \citet{grosse2023studying, akyurek2022towards} leverage influence functions and embedding similarities for item credit assessment. Additionally, other research \cite{wang2023wasa, sander2024watermarking} explores watermarking techniques for item tracking.

\section{Conclusion and Future Directions}\label{sec:future}

This survey has delved into the emergence of new bias and unfairness challenges within IR systems in this LLM era. 
We have established a unified framework to understand these issues as distribution mismatch problems and systematically categorized mitigation strategies into data sampling and distribution reconstruction approaches. Through an in-depth review of fifteen types of bias and unfairness, along with their corresponding mitigation strategies, we provide a comprehensive overview of the current progress. Despite the considerable attention given to this topic, we identify some important problems for further exploration.

\textbf{Biases and Unfairness in IR Feedback Loops.} In real IR systems, the interaction between users, models, and information forms feedback loops that impact each other over time. These loops can significantly shape user perceptions and preferences based on the information they are exposed to. This interaction, in turn, influences the training data, creating a cycle that may reinforce existing biases and unfairness. Novel strategies to interrupt the feedback loops are essential for mitigating these issues.

\textbf{Unified Mitigation Framework.}
Current methods primarily address individual instances of bias or unfairness, but in the future, we should consider unified solutions. This is because various types of bias and unfairness are not isolated. Presenting a unified framework can facilitate a deeper understanding of these relationships, enabling methods for addressing different types of bias and unfairness to complement each other. Our proposed unified perspective offers a potential direction to address these issues simultaneously.

\textbf{Theoretical Analysis and Guarantees.} The current exploration of bias and unfairness within the intersection between LLMs and IR systems has predominantly been through empirical studies. However, there is a critical need for robust theoretical analysis to augment these empirical findings. 
Future efforts should focus on developing more rigorous analytical frameworks.

\textbf{Better Benchmarks and Evaluation.}
Most benchmarks currently utilized to study bias and unfairness within simulated environments. There is a crucial need for collecting large-scale, real-world datasets to enhance the evaluations and broaden research horizons.
Additionally, as LLMs increasingly draw upon existing online data to train subsequent generations, dynamic benchmarks are needed to be constructed. Consequently, future work can focus on exploring a systematic evaluation protocol for different bias and unfairness issues.

\begin{acks}
This work was funded by the National Key R\&D Program of China (2023YFA1008704), the National Natural Science Foundation of China (No. 62377044, No.62276248), Beijing Key Laboratory of Big Data Management and Analysis Methods, Major Innovation \& Planning Interdisciplinary Platform for the  ``Double-First Class” Initiative, PCC@RUC, funds for building world-class universities (disciplines) of Renmin University of China, and the Youth Innovation Promotion Association CAS under Grants No.2023111.
\end{acks}

\clearpage

\bibliographystyle{ACM-Reference-Format}
\balance
\bibliography{ref}

\end{document}